# The physics of a walking robot

J. Güémez [a,1], M. Fiolhais [b,2]

[a] *Departamento de Física Aplicada*
*Universidad de Cantabria*
*E-39005 Santander, Spain*

[b] *Departamento de Física* and *Centro de Física Computacional*
*Universidade de Coimbra*
*P-3004-516 Coimbra, Portugal*

**Abstract**
The physics of walking is explored, using a toy as a concrete example and a 'toy model' applied to it. Besides the Newton's second law, the problem is also discussed from the thermodynamical perspective. Once the steady state (constant velocity) is achieved, we show that the internal energy of the toy is dissipated as heat in the surroundings.

In a previous publication [1] we discussed the physics of toys and showed how useful they can be in the classroom and in demonstrations. However, an accurate physical discussion of a toy is usually intricate, though sometimes a quantitative discussion is possible under a limited number of simplifying but justifiable assumptions.

In this paper we analyse the physics of a walking person [2] or of some moving ("walking") toys. Many physics textbooks refer to walking in two perspectives: either to illustrate the third law of mechanics, or to give an example of a static friction force in the direction of the centre-of-mass motion of the system. It is common to find pictures in books where the action/reaction pair, on the foot and on the ground, is represented, and pictures just with the forward static friction force on the foot to show the force related to the motion. Actually, this is only part of the story, since a force on the same foot, in the opposite direction, should also be present, otherwise the walker would experience a permanent forward acceleration.

Of course, besides a (time dependent) horizontal force, to be discussed below, there are two vertical forces applied to the system, namely the weight and the normal reaction, whose resultant is not exactly zero. The normal force is also time dependent and this leads to a vertical oscillation of the centre-of-mass around an average height with respect to the ground [3], but here we are not interested in this discussion. We are mainly interested in the discussion of the horizontal motion, therefore assuming that the weight and the normal force exactly cancel each other.

Shortly after the walking starts, a 'stationary regime' is achieved, and the walker velocity gets (approximately) constant. Actually, the velocity of a walking person is not strictly constant, but there are only small oscillations around an average value. This is the regime we want to discuss.

As mentioned above, the horizontal force acting upon the walker's foot is both in the forward and in the backward directions, in different time periods that repeat cyclically. The impulse of that force should be zero in a cycle (one cycle is one stride) to keep the linear momentum and, therefore, the velocity at a constant value. In a recent paper [4] the horizontal forces acting upon a walking and a running person were presented (real data) as a function of time. In order to better understand the real situation with a person, it is appropriate to discuss the problem in an idealized situation, for instance with a walking robot or toy. This is the goal of this paper.

---

[1] guemezj@unican.es
[2] tmanuel@teor.fis.uc.pt

Let us consider a toy with an internal source of energy, for instance elastic energy stored in a spiral spring (a common situation in many toys). When the spring is released, we assume that the energy decrease is given by

$$\Delta U = \frac{1}{2} k(\theta^2 - \theta_0^2) \tag{1}$$

with $U_0 = \frac{1}{2} k \theta_0^2$ the initial stored energy and $k$ the elastic constant characterizing the spring (the angle $\theta$, with initial value $\theta_0$, is a decreasing function of time). As this energy decreases, the articulated toy or robot moves step by step as figure 1 shows. Other sources of internal energy are also common, as in the battery operated toys. However, the nature of the internal source of energy is immaterial for our discussion.

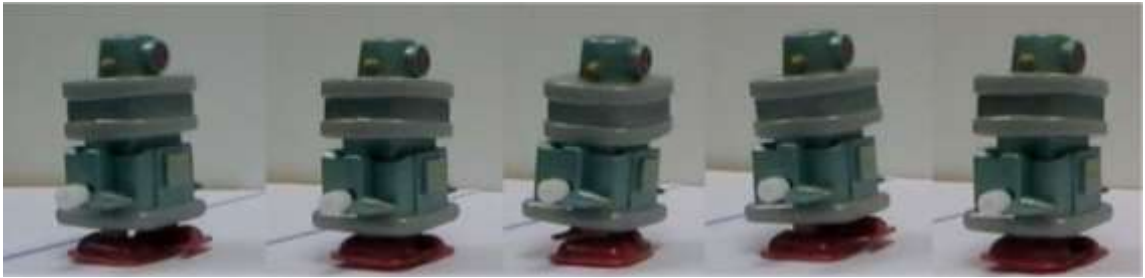

Figure 1: A walking toy.

Next, we analyse in some detail the horizontal force acting on the toy. A force applied by the ground on the foot in contact with the ground, in the forward direction, forces the centre-of-mass to accelerate in that direction. At some point the other foot contacts the ground and the resultant horizontal force is initially in the opposite direction, namely pointing backwards and forcing the centre-of-mass velocity to decrease. On the same foot (the other is now and still lifted), and after a while, the direction of the resultant force changes and again forces the centre-of-mass to increase its velocity. The process repeats regularly with a certain period. The force communicated by the ground is a continuous time dependent function as shown in [3, 4] for a walking person. The important point is that the impulse, in a complete cycle, be zero.

The impulse-momentum and the centre-of-mass energy equations for a system, whose differential forms are [5]

$$\vec{F}_{ext}\, dt = M\, d\vec{v}_{ext}, \quad \text{and} \quad \vec{F}_{ext} \cdot d\vec{r}_{cm} = \frac{1}{2} M\, dv_{cm}^2, \tag{2}$$

where $M$ is the mass and $v_{cm}$ the centre-of-mass velocity, apply to the motion of the toy or the person.

In order to keep the discussion at the simplest level, we may assume a 'toy model' for the horizontal force, applicable, so to say, to a robot: we suppose that it includes sections of constant magnitudes $F$ (in the forward direction) and $F'$ (in the opposite direction) in the time intervals $\Delta t$ and $\Delta t'$, respectively. Figure 2 represents such a horizontal force as a function of time. The actual values of $F$, $F'$ and of the time intervals, $\Delta t$, $\Delta t'$, do not matter for the discussion.

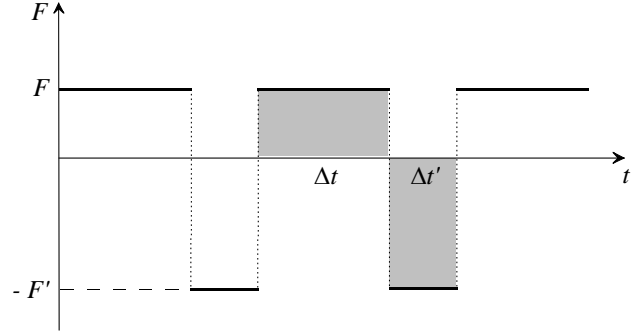

Figure 2: Resultant horizontal force acting upon the robot as a function of time.

A complete cycle (say, "one stride") corresponds to the time interval $\Delta t + \Delta t'$, during which the total impulse is zero. We could use more realistic forces, but the integration of equations (2) would then be more difficult. In our simplified model, that integration is straightforward, leading to

$$\begin{cases} M(v_\text{f} - v_\text{i}) = F\Delta t \\ \frac{1}{2}M(v_\text{f}^2 - v_\text{i}^2) = F\Delta x \end{cases} \quad (3)$$

where $v_\text{i,f}$ are the initial and final centre-of-mass velocities, and $\Delta x$ is the displacement of the centre-of-mass in the time interval $\Delta t$. Similarly, for the other phase of the step, one has,

$$\begin{cases} M(v'_\text{f} - v'_\text{i}) = -F'\Delta t' \\ \frac{1}{2}M(v'^2_\text{f} - v'^2_\text{i}) = -F'\Delta x' \end{cases} \quad (4)$$

where $v'_\text{i,f}$ are the initial and final centre-of-mass velocities in this phase of the step, and $\Delta x'$ the displacement of the center of mass, now in the time interval $\Delta t'$. The final velocity of this phase is equal to the initial velocity of the subsequent phase, $v'_\text{f} = v_\text{i}$ and, similarly, $v'_\text{i} = v_\text{f}$. Therefore, one immediately concludes that for the 'steady regime' walking

$$I = F\Delta t = F'\Delta t' = I' \quad \text{and} \quad W = F\Delta x = F'\Delta x' = W'. \quad (5)$$

The corresponding velocity of the centre-of-mass system is plotted in figure 3, as a function of time, where by $v_0$ we denote the "cruising" velocity. It is important to note that the friction forces are static ones and that the displacements in (5) are centre-of-mass displacements.

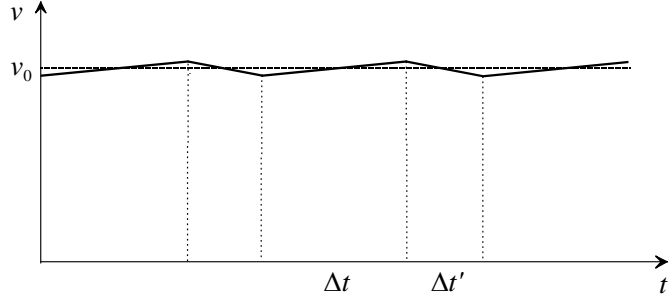

Figure 3: Velocity of the walking robot as a function of time.

The horizontal force as shown in figure 2 is useful to simplify the integration of the differential equations (2). Should we use a realistic force, the results expressed by (5) would be substantially the same: $I$ and $I'$ still equal in magnitude and $W$ and $W'$ also equal in magnitude, but now not simply given by products of the force by a time interval or by a displacement. Moreover, the edges in figures 3 would be smoothed out, corresponding to a continuous acceleration function. Nevertheless, and in spite of its simplicity, the used model accounts for the most fundamental issues of the walking dynamics. A more realistic force, inspired in the data presented in [4], would be of the type $F = F_0 \sin(\omega t)$, with $\omega$ the stride frequency. In this case $\Delta t = \Delta t' = \pi/\omega$ and the impulse in each semi-period is $I = 2F_0 \Delta t/\pi$. The velocity is $v = v_0 - \Delta v \cos(\omega t)$, where $\Delta v = F_0/(M\omega)$, whose graph, even quantitatively, is very close to the velocity represented in figure 3 (for $\Delta t = \Delta t'$). In a stride, as important as the forward force is the backward force in the same foot but in a different time interval. Walking is a succession of acceleration and deceleration but, for a cycle, there is no variation of the linear momentum and of the kinetic energy: $\Delta p_{cm}=0$ and $\Delta K_{cm}=0$, respectively.

Energetic issues are also important to be analysed and interpreted. Walking is possible if the walker — a person, a toy, a robot — possesses an internal source of energy. The not so uncommon idea that this energy transforms into kinetic energy is not true, simply because the kinetic energy is kept (approximately) constant. So, into what is that energy transformed? To answer the question we need to analyse the problem in the perspective of the first law of thermodynamics. The first law of thermodynamics is, in general, expressed by [5] $\Delta K_{cm} + \Delta U = W_{ext} + Q$, where, on the left-hand side, we identify the variation of the centre-of-mass kinetic energy and the variation of the system's internal energy (due to all possible reasons) and, on the right-hand side, we identify the work performed by the *external* forces, and the other possible energy transfers to/from the system that are not work – so they are heat. We could analyse each phase of the complete cycle, separately, but we can also analyse directly the full cycle, *i.e.* the complete step. On the left-hand side, since there is no variation of kinetic energy of the system as a whole, there is only variation of the internal energy, given by $\Delta U = k(\theta^2 - \theta_0^2) < 0$. On the right-hand side, there is no external work because the static friction forces do not do work [6], hence, there is only heat. Altogether, the first law of thermodynamics allows us to write

$$Q = \Delta U \qquad (6)$$

and this is energy dissipated in the surroundings [7]. Therefore one concludes that, during walking, there is a direct conversion of "organized" internal energy (the elastic energy stored in the spring) into "disorganized" energy in the surroundings. The entropy increase of the universe is $\Delta S_U = -Q/T > 0$, where $T$ is the temperature of the surroundings (assumed as a heat reservoir). These results hold irrespective of the detailed form of the time dependence of the static friction force — the variation of internal energy is totally dissipated as heat in the surroundings.

In the case of a real person, instead of a toy or a robot, the internal source of energy is provided by the biochemical reactions that take place in the person's muscles [8]. Now there is a work performed against the external atmospheric pressure and there is some heat associated with the reaction. The first law of thermodynamics, applied as in [5] for a person pushing against a wall, leads to $Q = \Delta G$, where $\Delta G < 0$ is the variation of the free Gibbs energy (and not simply the internal energy) associated with the biochemical reactions.

In the present study we assumed that the centre-of-mass was always at the same vertical level. We could easily go beyond this restriction by assuming a resultant vertical force also time dependent, with a certain period, and with a vanishing net vertical impulse. The mechanical treatment would be the same as for the horizontal motion. The behaviour of the vertical component of the centre-of-mass would be similar to figure 3, now with $v_0 = 0$. Regarding the thermodynamical conclusion expressed by equation (6) it also would not be changed if the vertical degree of freedom would be included in our treatment.

From the energetic point of view, walking and the motion of a car [9] have a lot of features in common. Static friction forces on the wheels or on the feet do not do any work. On the other hand, the energy, initially stored as chemical energy, in the person's muscles or in the fuel (and the air), is exhausted as heat or as dissipative work.

We analysed the dynamics and also the thermodynamics of walking, using a simplified model applied to a robot. In spite of its simplicity the model accounts for the most significant aspects of the phenomenon. A more refined model would not change the qualitative view. In particular we showed that the internal energy variation of the moving body is exhausted as heat dissipated in the surroundings and this is a model independent result.